\documentclass[12pt,a4paper,oneside]{article}

\usepackage{graphicx}
\usepackage{overcite}
\usepackage{amsmath}

\usepackage[sort&compress,numbers,super]{natbib}

\newcommand{\ket}[1]{\ensuremath{| \: #1 \: \rangle}}

\title{\textsc{autoCAS}: a program for fully automated multi-configurational calculations}
\date{May 07, 2019}
\author{\vspace{1cm}Christopher J. Stein and Markus Reiher\footnote{Corresponding author: markus.reiher@phys.chem.ethz.ch}\\
\textit{ETH Z\"urich, Laboratorium f\"ur Physikalische Chemie,}\\ 
\textit{Vladimir-Prelog-Weg 2, 8093 Z\"urich, Switzerland}}

\begin{document}

\maketitle

\begin{abstract}
We present our implementation \textsc{autoCAS} for fully automated multi-configurational calculations, which we also make available free of charge on our webpages.
The graphical user interface of \textsc{autoCAS}  connects a general electronic structure program with a density matrix renormalization group program to carry out our recently introduced automated active space selection protocol for multi-configurational calculations [\textit{J. Chem. Theory Comput.}, \textbf{2016}, \textit{12}, 1760].
Next to this active space selection, \textsc{autoCAS} carries out several steps of multi-configurational calculations so that only a minimal input is required to start them, comparable to that of a standard Kohn-Sham density functional theory calculation, so that black-box multi-configurational calculations become feasible.
Furthermore, we introduce a new extension to the selection algorithm that facilitates automated selections for molecules with large valence orbital spaces consisting of several hundred orbitals.

\end{abstract}

\newpage

\section{Introduction}

The correct quantum chemical description of numerous chemically relevant systems and processes, such as many transition-metal complexes in ground- and excited states, molecules with extended $\pi$-systems, or bond-breaking and bond-forming processes, demands more than one electronic configuration.
Although multi-configurational methods exist that attempt to directly incorporate all important electronic configurations,\cite{huro73,cimi87,mira93,tubm16,holm16,boot09,boot10,shep12,zimm17,erik17,erik18,ivan01,ivan02} methods based on the complete active-space self-consistent field concept (CASSCF) \cite{roos80,rued82,rued82a,rued82b,wern85,roos87,shep87} are the \textit{de facto} standard for multi-configurational calculations with applications ranging from organic photochemistry\cite{andr05} to inorganic catalysis.\cite{bern16}
In these methods, the crucial task of selecting the configurations with a large weight in the final wave function is transferred to a selection of a subset of orbitals --- and corresponding electrons --- for which all possible configurations, including the most important ones, are constructed.
A CASSCF wave function is essentially a full configuration interaction wave function in the pre-defined orbital subspace, usually referred to as the \textit{active space}.

The proper selection of this active space is crucial to obtain reliable multi-configurational wave functions and corresponding energies.
Unfortunately, this delicate selection is mostly achieved by tedious trial-and-error procedures,\cite{pier01,oliv15,ashl17} several empirical guidelines,\cite{roos87,pier01,pier03,very11} and chemical intuition,\cite{angl95,pier01,kerr15,ashl17} a term that lacks a clear definition in this context.
Obviously, this is a not at all satisfactory situation for a highly advanced \textit{ab initio} method.
Attempts to settle this selection on more solid physical ground were mostly based on natural orbital occupation numbers (NOONs) of approximate wave functions that may help to identify strongly entangled orbitals that must be included in the active space.
The NOONs of unrestricted Hartree--Fock,\cite{bofi89,kell15} second-order M{\o}ller--Plesset perturbation theory\cite{jens88}, partially converged density matrix renormalization group (DMRG)\cite{wout14a}, or $N$-electron valence state theory wave functions\cite{khed19} were proposed as suitable approximate wave functions.
For all these approaches, however, the threshold, that defines when the deviation of a NOON from 0 or 2 is large enough to incorporate the corresponding orbital into the active space, cannot be easily set for arbitrarily complex electronic structures
(picture the dense orbital region around the Fermi level in a polynuclear anti-ferromagnetically coupled transition-metal cluster 
where many valence orbitals exhibit gradually changing NOONs in the intervals of, say,  [0,0.35] and [1.65,2.0]).
Hence, while making the approach more transferable than a manual active-space selection, none of these methods has enabled fully automated multi-configurational calculations and a high level of expertise has still been required.

In 2016, we proposed to base the active space selection on orbital entanglement entropies calculated from partially converged DMRG calculations.\cite{stei16}
We adapted a single threshold for the entropy measure to the overall static correlation of the wave function, which allowed us to define a transferable threshold value.
With this definition, we could fully automate the active space selection procedure as there remained no system-dependent parameter.
Very recently, Legeza and coworkers also reported an active orbital space selection based on orbital entanglement entropies\cite{faul19} but they do not detail the selection procedure and the choice of a suitable cutoff.

In 2017, Chan and coworkers proposed an elegant automated construction of valence active spaces\cite{sayf17} from intrinsic atomic orbitals.\cite{kniz13}
These valence active spaces are usually too large to be included in multi-configurational calculations but the authors propose a method to truncate these valence active spaces based on the eigenvalues of a projected overlap matrix generated during the construction of these orbitals.
Alternatively, one might truncate these valence active spaces with our 2016 active space selection protocol detailed below.
In a recent study this approach was extended to an automated construction of $\pi$-orbital active spaces for conjugated systems.\cite{sayf19}

A semi-automated active space selection for rather small molecules was proposed by Bao \textit{et al}.\cite{bao18}
This three-step/three-parameter scheme was developed to automate the calculation of excitation energies of small doublet radicals with up to 15 electrons.
An extension of this scheme to larger systems and more demanding processes such as bond-breaking has not been presented yet and its general applicability still needs to be demonstrated.

A driving force of our developments was the desire to develop an automated selection of the active orbitals, a suitable choice of other computational parameters, and an overall simplified setup to steer multi-configurational calculations in a black-box fashion, which would make them as easily accessible as standard Kohn-Sham density functional theory\cite{kohn65,parr94} (KS-DFT) calculations.
Apart from the fact that this makes multi-configurational calculations more accessible (a practical perspective) as well as unbiased,  
objective, and rigorous (a conceptual perspective), we foresee an important future of these ideas in the light of data-driven computational chemistry, for which the automated benchmarking and uncertainty quantification\cite{Proppe2017a,Simm2018,Proppe2019} in automated reaction network explorations\cite{simm19} is one example.
Therefore, the \textsc{autoCAS} program\cite{autocas} that we present in this work has been designed to meet i) the desire to turn multi-configurational approaches into black-box methods as well as ii) the requirements of experienced experts and iii) those of novice computational chemists. 

We first provide a brief review of the underlying orbital entanglement based automated active space selection and describe the details of the implementation. We then introduce an extension that allows for an automated active space selection for molecules with large valence orbital spaces.
The final sections of this work describe recommended settings and future work, before we summarize the results.

\section{Theoretical background}
We presented the automated active space selection protocol in detail in Ref.~\citenum{stei16} and will only review the most important underlying principles here.
In Ref.~\citenum{stei16a} we further demonstrated that the automatically selected active spaces are not only more compact than manually selected active spaces\cite{phun12} based on empirical rules,\cite{roos87,pier01,pier03,very11} but are also well suited for subsequent perturbation theory calculations.
Extensions to the automated active space selection for excited states and reaction coordinates are described in Ref.~\citenum{stei17}.

\subsection{The density matrix renormalization group}
The number of configurations that can be constructed by distributing a given number of active electrons $N$ among the active orbitals $L$ grows exponentially, although spin and spatial symmetry reduce this number drastically.\cite{aqui16}
It was realized quite early\cite{ivan01,ivan02} that a large number of these configurations have a negligible weight in the optimized wave function and hence a negligible contribution to the energy.
Several computational models were devised to separate the configurational "deadwood" from those configurations that are essential for the qualitatively correct description of the multi-configurational wave function.
Among those are selected CI approaches\cite{huro73,cimi87,mira93,nees03,tubm16,holm16} and full configuration interaction quantum Monte Carlo.\cite{boot09,boot10,shep12}
A third approach that effectively reduces the number of configurations in the CI expansion is DMRG.\cite{whit92,whit93,lege08,chan08,chan09,mart10,mart11,chan11,scho11,kura14,wout14,yana15,szal15,knec16,chan16,baia19}
DMRG iteratively optimizes a wave function that is expressed as a matrix product state\cite{romm95} (MPS)

\begin{align}
  \ket{\Psi_k} = \sum_{\sigma_1,...,\sigma_L} \sum_{a_1,...,a_\text{L-1}}^m 
               {M}_{1,a_1}^{(k)\sigma_1} {M}_{a_1,a_2}^{(k)\sigma_2}...
               {M}_{a_\text{L-1},1}^{(k)\sigma_L}
               \ket{\sigma_1,...,\sigma_L}\, .
               \label{mps}
\end{align}

In this equation, a matrix $M^{(k)\sigma_i}_{a_{i-1},a_i}$ is associated with each active orbital $i$, $\sigma_i = \{0, \uparrow, \downarrow, \uparrow \downarrow\}$ denotes the occupation of spatial orbital $i$, $m$ is the number of renormalized block states and $\ket{\sigma_1,...,\sigma_L}$ is the occupation number vector.
The fraction of the CI space recovered is controlled by $m$ and an optimal balance between cost and accuracy is aimed for in actual calculations.
Chemical accuracy for relative energies may be obtained with a moderate number of renormalized block states so that active spaces with up to 100 active orbitals can be handled by DMRG.
This is contrasted by about 18-24 orbitals in traditional CASSCF calculations.\cite{vogi17}
It is the fact that very large active spaces can be calculated in an iterative procedure that enables the automated selection of active spaces as will be detailed below.
We emphasize that crucial parameters are the number of macroiterations in the MPS optimization (called "sweeps" in the context of DMRG), the bond dimension $m$, the order in which the orbitals are arranged on a one-dimensional lattice for the sweeping, and the initial guess for the matrix elements in Eq.~(\ref{mps}).

\subsection{Orbital entanglement}
With DMRG it is possible to calculate a qualitatively correct wave function for a large initial active space.
Measures that quantify the importance of a given orbital for the description of static correlation may be calculated from this initial wave function.
The single-orbital entropy $s_i(1)$ is such a measure.\cite{lege03,riss06,lege06}
As a von-Neumann-type entropy,\cite{neum27} it quantifies the deviation from a pure state measured by the eigenvalues $w_{\alpha,i}$ of a grand-canonical one-orbital reduced density matrix (1o-RDM)
\begin{align}
s_i(1) = - \sum_{\alpha = 1}^4 w_{\alpha,i} \ln w_{\alpha,i} \, ,
\end{align}
where the 4 states of each spatial orbital $i$ are unoccupied, doubly occupied, spin-up and spin-down.
Note that the 1o-RDM is already diagonal because of particle number and spin conservation.
Analogously, two-orbital entropies can be defined from the two-orbital reduced density matrix (2o-RDM)
\begin{align}
s_{ij}(2) = - \sum_{\alpha = 1}^{16} w_{\alpha,ij} \ln w_{\alpha,ij} \, ,
\end{align}
where $w_{\alpha,ij}$ are the eigenvalues of the 2o-RDM and all 16 combinations for the occupation of two orbitals  $i$ and $j$ are considered.
The two-orbital entropy of an orbital pair describes by how much the occupation of states defined on this orbital pair are affected by the presence of all other orbitals.
The one- and two-orbital entropies can be combined to give the mutual information which describes how two orbitals influence each others occupation
\begin{align}
I_{ij} = \frac{1}{2}[s_i(1) + s_j(1) - s_{ij}(2)](1-\delta_{ij})\, .
\label{eq:mutual}
\end{align}

The single-orbital entropies and the mutual information evaluated for a given wave function can be displayed in an \textit{entanglement diagram} (see Fig.~1 for a simple example).
\begin{figure}[t!]
\begin{center}
\includegraphics[width=\textwidth]{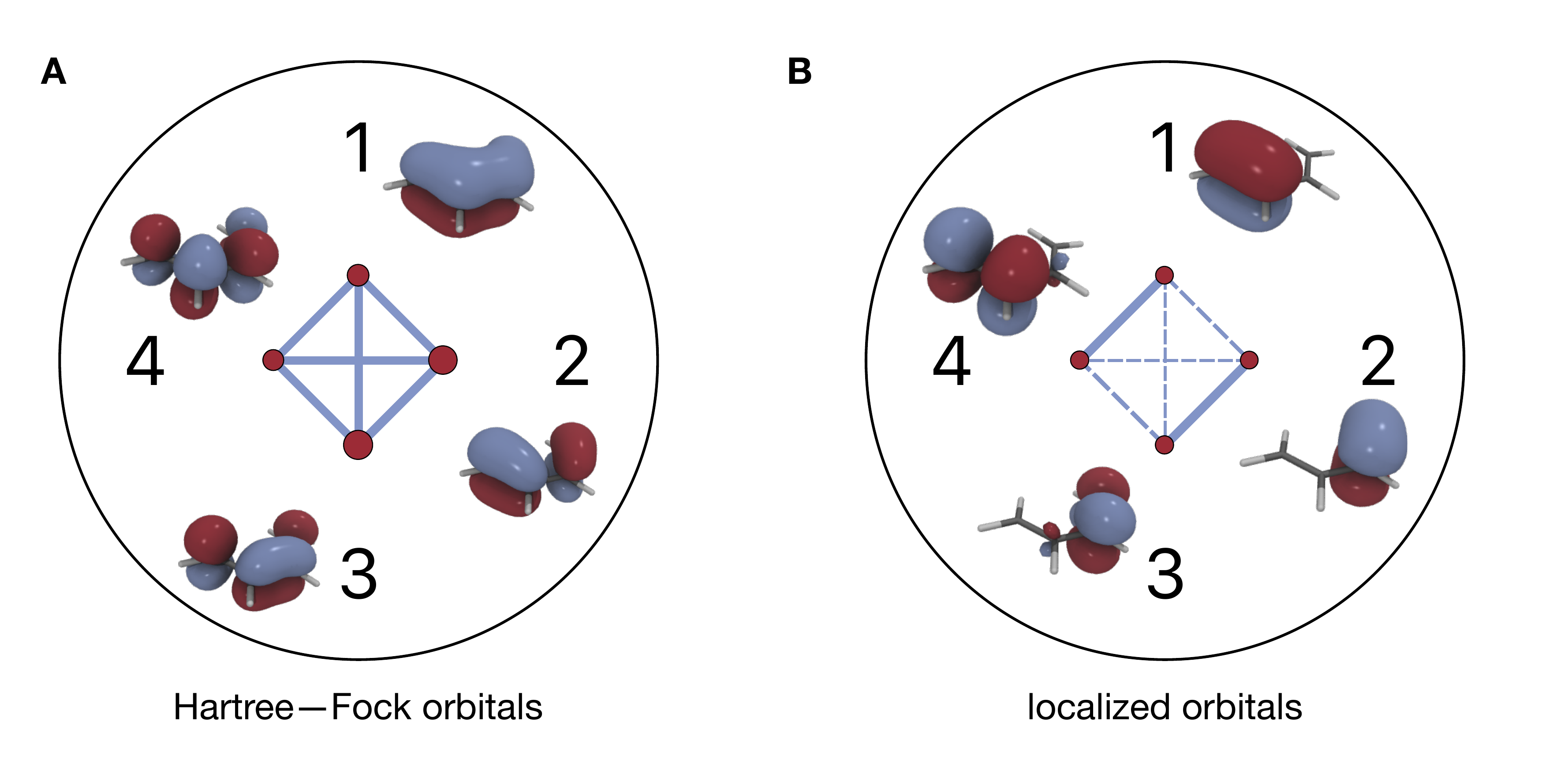}
\caption{Entanglement diagrams for a converged calculation of the $\pi$-orbital active space of \textit{cis}-1,3-butadiene in the basis of canonical Hartree--Fock orbitals (panel A) and localized orbitals (panel B). All orbitals are mutually entangled in the canonical orbital basis, whereas a pronounced $\pi$/$\pi^*$ entanglement pattern can be identified for the localized orbitals.}
\label{entanglement}
\end{center}
\end{figure}
In these diagrams a circle is assigned to each orbital of the active space.
The area of this circle is proportional to the single-orbital entropy of this orbital.
The circles themselves are connected by lines whose thickness and color intensity reflect the value of the mutual information of the orbital pair.
These diagrams allow one to extract plenty of information about the wave function and the underlying orbital basis.
We note, however, that the orbital entanglement measures introduced above depend on the orbital basis and do not provide unambiguous information on the multi-configurational wave function.
In cases where a detailed wave function analysis with transferable results is desired,\cite{mai18} it is advisable to carry out the entanglement analysis in a well-defined orbital basis such as natural orbitals.
In the context of the automated active space selection, however, this basis set dependence of the orbital entanglement is beneficial for two reasons:
\begin{enumerate}
\item The starting orbitals can be chosen arbitrarily without any effect on the selection algorithm. While full automation might call for simple Hartree--Fock orbitals as the orbital basis, computational or time constraints call for a manual preselection of orbitals which is easiest in a localized orbital basis that reflects bonding and anti-bonding orbitals. Since the entanglement entropy measures can be evaluated for any orbital basis, full flexibility is achieved and the automated selection algorithm does not need to be adjusted.
\item It can be beneficial to switch between orbital bases during the orbital optimization process. While initial DMRG calculations with many orbitals might require a localized orbital basis, because DMRG has been shown to optimize wave functions in such a basis most efficiently,\cite{mitr12} subsequent calculations with orbital optimization may be carried out in a self-consistent DMRG-SCF or CASSCF orbital basis.
\end{enumerate}

In the automated active space selection, we exploit the single-orbital entropy (and, if deemed necessary, also the mutual information) to rank the contribution of an orbital to the static correlation in a wave function.
Static correlation is present when there are several configurations with a large weight in the total wave function.
This directly translates to a high single-orbital entropy for orbitals with different occupation in such configurations.
While the ranking of the orbitals according to their single-orbital entropy is straightforward, it is important to define a cutoff threshold that clearly defines which orbitals to include and which to reject.
A transferable threshold can be identified by scaling the individual single-orbital entropies to the largest single-orbital entropy of a given calculation.
The exact algorithm for the cutoff selection is described in Ref.~\citenum{stei16} and we note here that the definition of the cutoff is the only parameter in our automated selection algorithm.
It is set to a standard value of 10\% in \textsc{autoCAS},  which proved to be a reliable and rather universal definition.
The selection threshold was carefully adjusted\cite{stei16} to reproduce previously reported active orbital spaces that were shown to give reliable total energies when dynamical correlation was included. We confirmed the reliability of this threshold in several subsequent studies.\cite{sinh17,stei17,stei17a} The automatically selected active spaces were either exactly those proposed in earlier studies or they were more compact but yielded the same accuracy when dynamical-correlation corrections were added.

An additional cutoff based on the multi-configurational diagnostic $Z_{s(1)} $ (\textit{vide infra}) signals single-configurational wave functions but is not essential to the algorithm.
Our orbital selection criterion has, so far, been solely based on the single-orbital entropy and has not required additional information from the mutual information.
We investigated in detail if alternative selection criteria that take the mutual information into account lead to other and more reliable orbital selections but found that not to be the case for the systems studied.\cite{stei16,stei16a,sinh17,stei17,stei17a}
Hence, in the spirit of Occam's Razor, we decided to base our selection only on the single-orbital entropies until a case will be discovered for which the mutual information provides additional critical information.

As noted above, alternative selection criteria have been proposed that are based on the natural orbital occupation numbers (NOONs) of an unrestricted Hartree--Fock calculation,\cite{bofi89,kell15} M{\o}ller-Plesset perturbation theory calculations,\cite{jens88} and various other wave-function models.\cite{krau14,wout14a}
While our automated selection method does not contradict any active space composition proposed by these methods, we find our approach to be more general and stable.

As a last remark we note that the orbital entanglement measures can be obtained from only partially converged wave functions because the accuracy required here (typically only about two decimals) is much lower than for the energy, where micro-Hartree accuracy is usually desired.
As a consequence, a qualitatively correct, but energy-wise not fully converged DMRG wave function serves the purpose.
Note that it is straightforward to detect qualitative deviations of the entanglement entropies of a final small-CAS fully converged DMRG calculation from a loosely optimized small-$m$ large-CAS one, so that inconsistencies can easily be spotted and cured.

\subsection{Multi-configurational diagnostic $Z_{s(1)}$}

Since the orbital entanglement measures are directly related to the static correlation they are a suitable basis for the definition of a multi-configurational diagnostic.
We defined such a diagnostic\cite{stei17a} as a normalized sum over the single-orbital entropies of the orbitals that are selected by the automated selection protocol
\begin{align}
Z_{s(1)} = \frac{1}{L \ln 4} \sum^L_i s_i(1) \, ,
\label{diagnostic}
\end{align}
where $L$ is a subset of orbitals included in the initial calculation.
The subset is defined by including only those orbitals that are selected by the automated protocol and the additional constraint that the number of orbitals must equal the number of electrons $N$ so that maximum entanglement can in principle be achieved.
The normalization factor is the maximum entanglement for a multi-configurational wave function under the afore-mentioned constraints.

\textsc{autoCAS} evaluates the diagnostic automatically for a given selection of active orbitals.
In fact, the diagnostic is even included in the selection protocol in two ways:
First, a warning is issued for $0.1< Z_{s(1)}<0.2$ stating that the multi-configurational character of the wave function is rather low and single-configurational methods might give more accurate results at lower computational cost.
Second, if the automated selected active space results in a $Z_{s(1)}$ value that is lower than 0.1, a smaller active space is selected until $Z_{s(1)}>0.1$.
This guarantees that for wave functions with low multi-configurational character no mainly dynamically correlated orbitals are included because those are likely to rotate out during the orbital optimization and size-consistency cannot be achieved.
The $Z_{s(1)}$ diagnostic may therefore replace the determination of the overall multi-configurational character based on the largest single-orbital entropy that was proposed in our initial implementation.\cite{stei16}
Note, however, that $Z_{s(1)}$ is only well-defined for active spaces selected with the automated selection protocol.
Hence, although \textsc{autoCAS} will display a $Z_{s(1)}$ value also for manually selected active spaces, it is important to note that the diagnostic cannot be compared between different calculations with different active spaces.

\section{Technical details}

\begin{figure}[t!]
\begin{center}
\includegraphics[width=\textwidth]{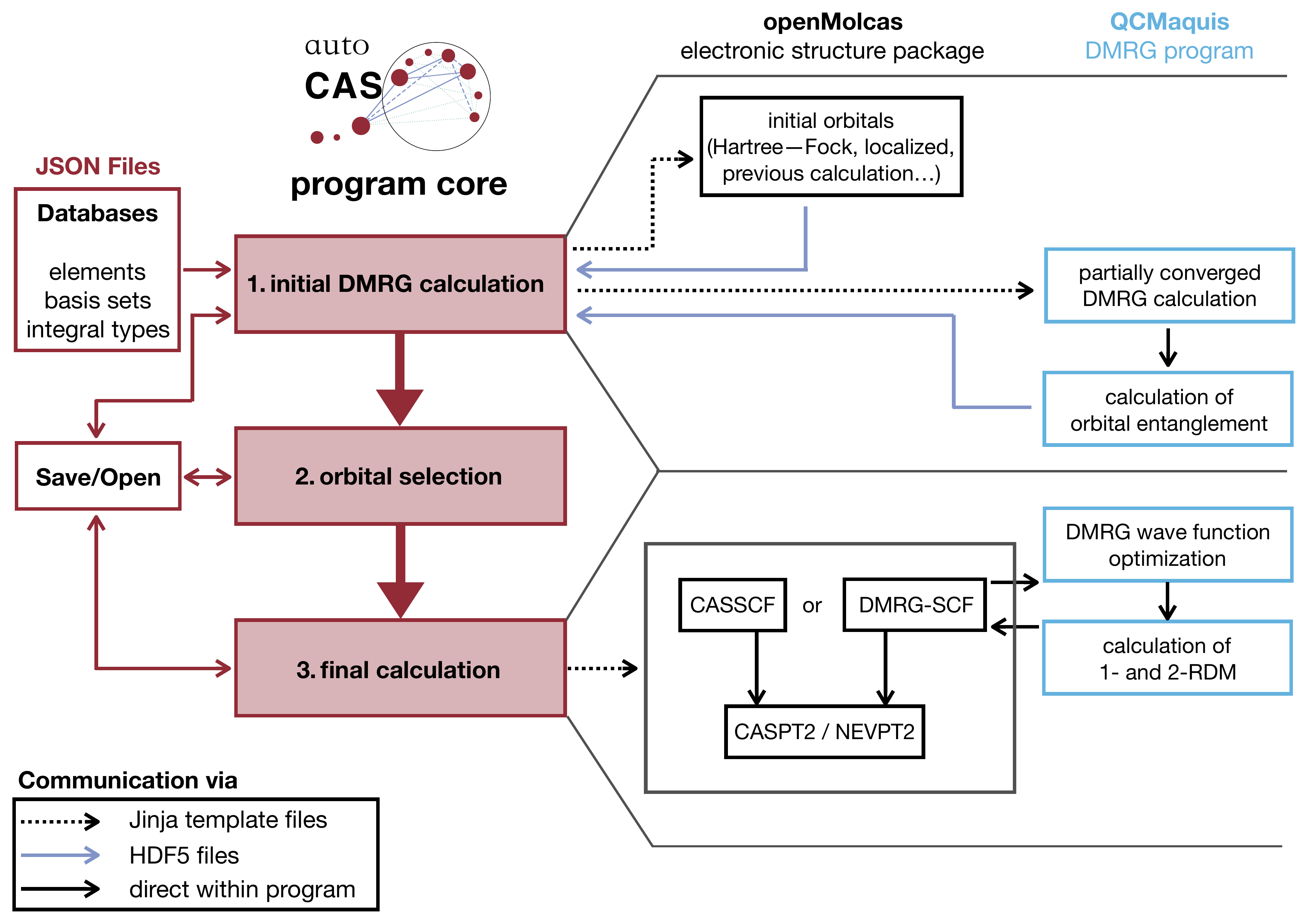}
\caption{Scheme of the program core and corresponding communication structures of \textsc{autoCAS} interfacing the \textsc{openMolcas} electronic structure program and the \textsc{QCMaquis} DMRG program. The three main steps of the algorithm are highlighted in red. Inputs for the external programs are provide with Jinja template files, whereas communication of the results to \textsc{autoCAS} is realized with HDF5 files. Internal communication within \textsc{autoCAS} is carried out through JSON data structure files. }
\label{procedure}
\end{center}
\end{figure}

Because of the numerous applications, the automation of multi-configurational calculations is a desirable, but challenging task.
In addition to that, the multitude of hardware setups and usage scenarios complicates the automation of these calculations.
Furthermore, being the interface between a DMRG program and a general multi-purpose electronic structure program, \textsc{autoCAS} has to be designed such that future changes in any of the programs it controls can be easily accounted for.
Therefore, it is beneficial to define a program core that can only be marginally adjusted and is unlikely to be altered in future updates.
This also increases backward compatibility.
The program core of \textsc{autoCAS} consists of the general three steps of the active space selection (see Figure 2) with the selection protocol detailed in Ref.~\citenum{stei16} being the central element.

In order to minimize the manual input and to make the individual steps of the calculation transparent, \textsc{autoCAS} is a graphical user interface (GUI) written in the Qt framework.\cite{qt}
The active space for a molecule with given charge and spin state can be determined fully automatically or with manual control over every step in the calculation.
Naturally, the second approach allows one to modify the setup of the individual calculations to a greater extent than the first variant.
Internal databases about element specific information, basis sets and integral types are provided to \textsc{autoCAS} by means of JSON files.
This file format is also chosen to store the current status of a calculation to continue at a later time or in order to transfer calculations with high hardware requirements to suitable machines.
Since JSON files can be easily modified, these files might be supplemented with results from other programs if, for example, only the plotting capabilities of \textsc{autoCAS} are desired to process entanglement outputs obtained by other programs.
This might be of interest as the orbital entanglement can be calculated for any multi-configurational method.
The JSON project files that \textsc{autoCAS} produces in order to save and load a calculation can be supplied in the supporting information of a publication to substantially increase the reproducibility of these multi-configurational calculations.

In the current version, \textsc{autoCAS} interfaces the \textsc{openMolcas} electronic structure program\cite{aqui16} and the \textsc{QCMaquis} DMRG program,\cite{kell15a,kell16} which are both open-source software projects downloadable from the URLs in Refs.~\citenum{qcmaquis,openmolcas}.
Input files for these programs are generated from predefined Jinja templates.
These input files can be manipulated while \textsc{autoCAS} guides the practitioner through the several steps of the calculation but they may also be modified outside of the program in order to adjust the calculations to the requirements of the current study. 
Output information such as orbital files or the entanglement information is transferred to \textsc{autoCAS} through HDF5 files.\cite{hdf5}
The manipulated orbital files (after the automated selection process) are also supplied to \textsc{openMolcas} in this file format.
The entanglement diagrams and the threshold diagrams from which the selection threshold is determined can be directly exported as publication ready pdf files.

A suitable active space can automatically be selected for the electronic ground state or a given number of excited states.
In the latter case, the DMRG wave functions of the excited states are optimized in a ground-state search in the space orthogonal to the lower-lying electronic states.\cite{scho11}
Active orbitals are then selected for each state separately and the active space for the final calculation is constructed from the union of all selected orbitals.

Although the automated active space selection is the key feature of \textsc{autoCAS}, it is also possible to select active spaces manually by visual inspection of the entanglement.
In such a case, the calculation is still greatly simplified because all file transfer and input generation is carried out by \textsc{autoCAS}.
This is of high value when the active space convergence is studied or other reasons exist that call for specific active spaces that are not selected based on the entanglement information, a typical example being the reproduction of previously published data.

Optimization of the selected active orbitals and inclusion of dynamical correlation marks the final step of the calculation (see Fig.~2). 
The orbitals can be optimized with either CASSCF or DMRG-SCF and the latter method is automatically chosen by \textsc{autoCAS} for active spaces with more than 14 orbitals.
Two types of multi-reference perturbation theories are available for the calculation of dynamic correlation: second-order perturbation theory with a complete active-space
self-consistent field reference wave function (CASPT2)\cite{ande90,ande92,knec19} and $N$-electron valence state perturbation theory of second order (NEVPT2).\cite{ange01,ange01a,ange02,frei16}
Large atomic orbital basis sets can be selected by exploiting Cholesky decomposition of the two-electron integrals.\cite{beeb77,aqui07,aqui08,frei16}

Figure 3 shows the main screen of \textsc{autoCAS} after a successful calculation in panel A and the Project Designer that serves to set up the calculation in panel B. The main screen is dominated by the entanglement diagram of the current calculation. The occupation numbers and single-orbital entropies are displayed when hovering over the circles associated with each orbital. When the 'manual selection' mode is switched on, orbitals can be selected for the final active space with a click on that circle. The automated active space selection is invoked by a click on the 'Automated Selection' button. Once an active space has been selected, the corresponding $Z_\mathrm{s(1)}$ value is also displayed. Log messages are displayed in the lower part of the main screen in order to show the history of previous calculations and the status of the current calculation. The Project Designer (panel B in Figure 3) is the starting point for each calculation where project-dependent paths are set and details of the calculation such as charge, spin, basis set and the number of electronic states to be calculated are adjusted. Upon start of the GUI and during several steps of the calculations, pop-up messages inform about possible next steps and finished calculations and warn if a calculation failed. These messages are intended to guide the less experienced practitioner, but they can be switched off if deemed unnecessary. In addition, we provide a detailed manual\cite{autocas} that describes every button and field of all screens and widgets of the GUI and includes several step-by-step example calculations. Hints on how to customize the provided template files and an overview of possible usage scenarios are also included.

\begin{figure}[t!]
\begin{center}
\includegraphics[width=\linewidth]{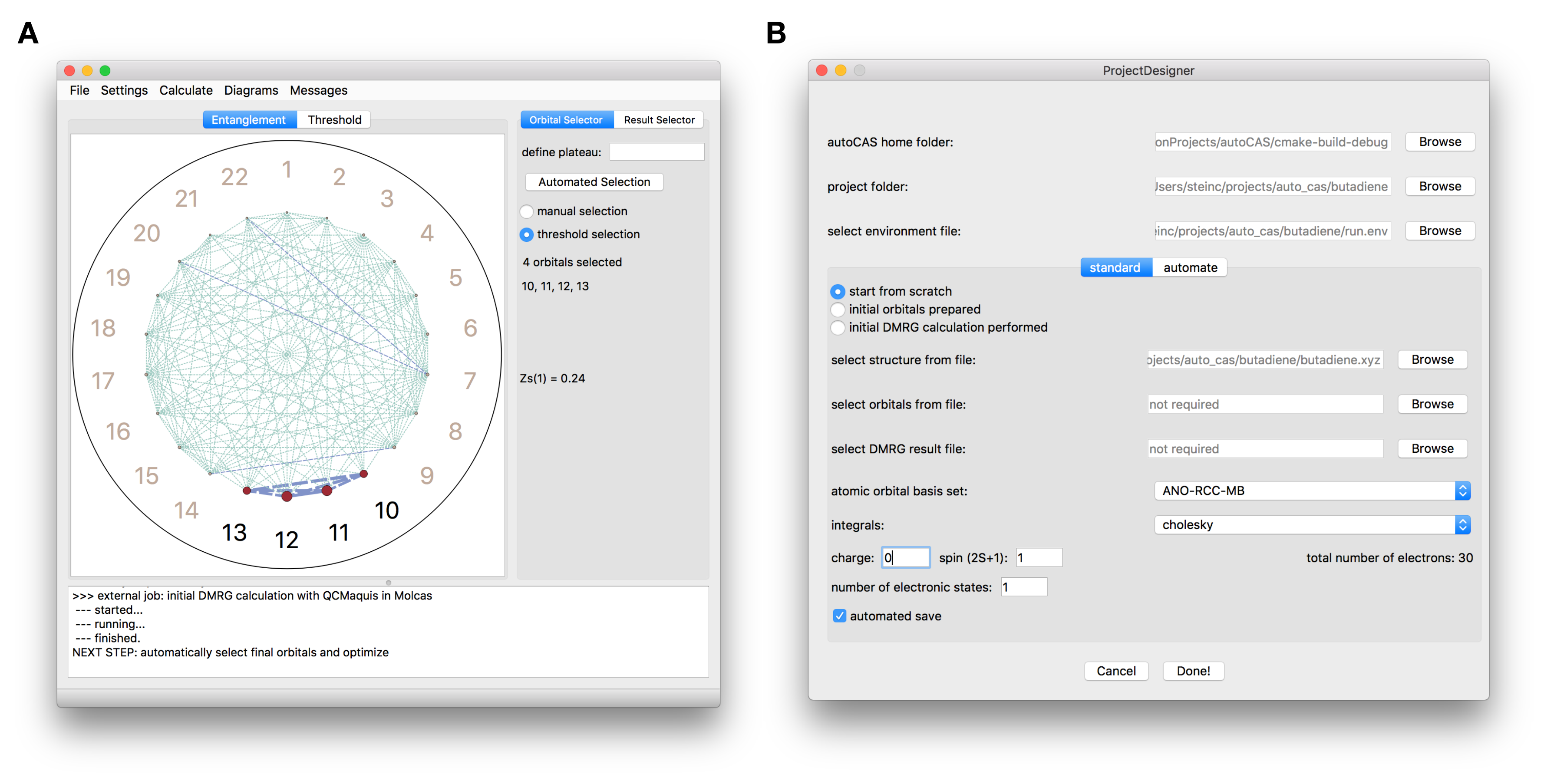}
\caption{Panel A: Main screen of \textsc{autoCAS} after a successful calculation on butadiene. The full $\pi$-space (orbitals 10-13) was automatically selected. Panel B: Project Designer widget for the setup of the calculations. A detailed explanation of all buttons and fields for all screens and widgets of \textsc{autoCAS} can be found in the manual.}
\end{center}
\end{figure}

\section{Ground-state calculations with very large valence spaces}

Although up to about 100 candidate orbitals can be investigated in an initial exploratory DMRG calculation, this certainly imposes an upper limit on the automated selection protocol.
Usually, this limit is circumvented by preselecting orbitals either by manual inspection or by imposing constraints such as maximum distance from a reactive center.
The manual effort of such a preselection is small compared to the tedious selection of active orbitals for a converged CASSCF-type calculation but any manual step certainly contradicts the spirit of a fully automated approach.
For a fully automated approach an unbiased candidate orbital preselection needs to be combined with an (approximate) evaluation of the entanglement entropies for a number of orbitals that is potentially much larger than what can be calculated in a standard DMRG calculation.
The full valence space is such an unbiased orbital subset that can be determined without manual input solely from information about the individual atoms in the molecule.
It is clear that Rydberg states or other highly excited states cannot be constructed from active spaces that contain only the valence orbitals and extensions will be needed for them.
However, for most common applications of multi-configurational calculations, the final active orbitals are recruited from the full valence space with the exception of $3d$-transition metals for which a second $d$-shell is often included to account for the double-shell effect.\cite{ande92}
Such an extension can, of course, easily be incorporateded in a fully automated approach.

The full-valence space, however, rapidly exceed the limit of about 100 orbitals on DMRG calculations as it scales with system size.
Therefore, a method must be devised that is capable to calculate the entanglement information for several hundreds of orbitals, at least approximately.
In a localized basis, a given orbital is often connected to only one or at most to a few other orbitals by a strong mutual information element.
This is because the interactions that cause static correlation are usually rather local (e.g. $\pi/\pi^*$-interactions, $\sigma/\sigma^*$-interactions in bond-breaking processes, or a distribution of the $d$-electrons among the $d$-orbitals in transition metal compounds).
It is therefore possible to split a large active space into smaller subsets and analyze the entanglement entropies of these subsets in order to identify these pairwise interactions. 
\begin{figure}[t!]
\begin{center}
\includegraphics[width=\textwidth]{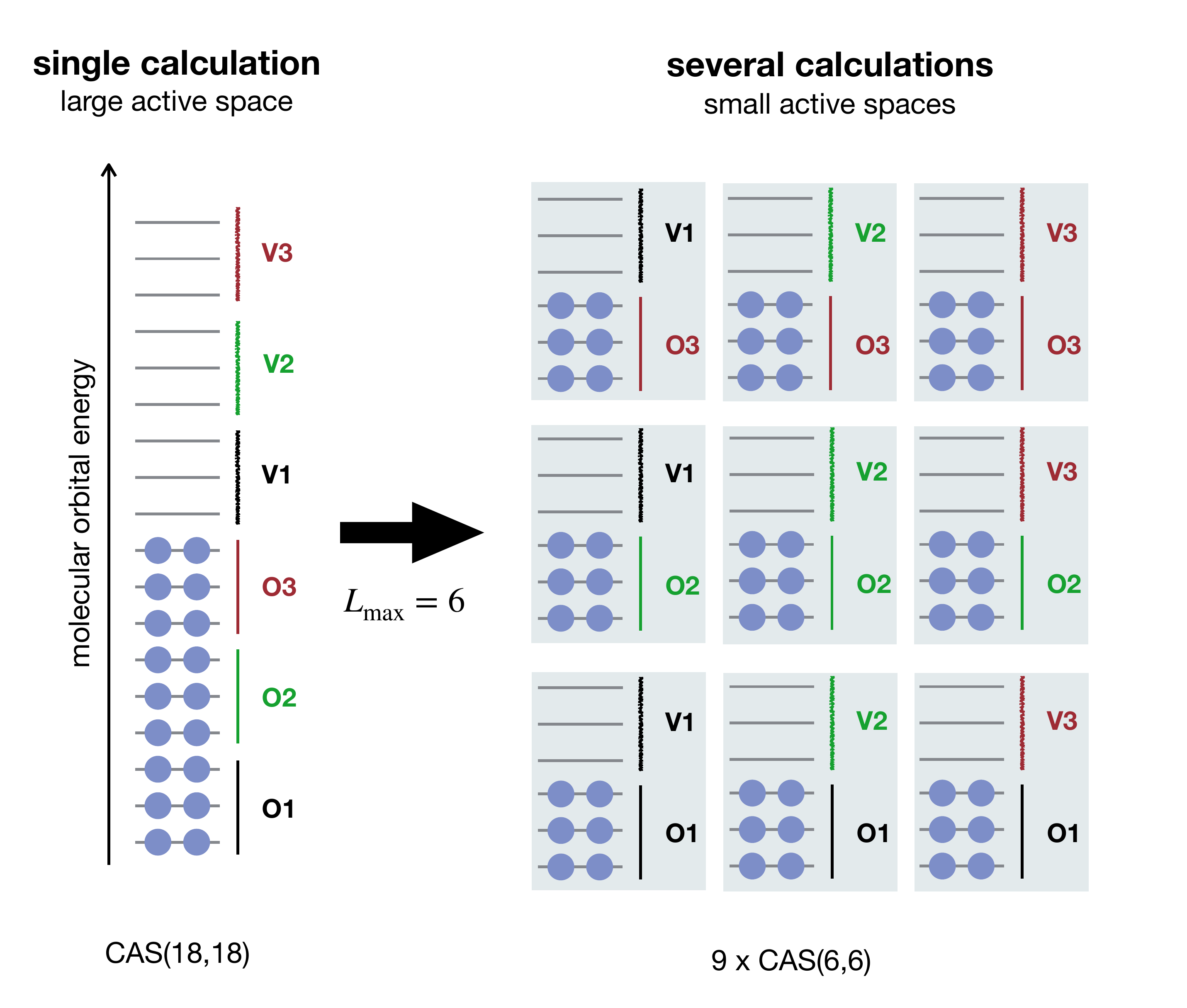}
\caption{Simple scheme for the overlapping dissection of a prohibitively large initial active space (such as the full valence space) into several smaller ones by splitting the occupied and virtual spaces into equal sized subspaces. Recombination of these subspaces leads to multiple distinct small active spaces with $L_\mathrm{max}$ orbitals. In the particular example shown here a CAS(18,18) calculation is split into nine CAS(6,6) calculations.}
\label{distribute}
\end{center}
\end{figure}
We introduce such a scheme in Fig.~\ref{distribute}, where a calculation for a CAS(18,18) is split into nine CAS(6,6) calculations.
The occupied and virtual spaces are separately split into subspaces and recombined to yield several smaller active spaces of size $L_\mathrm{max}$, which becomes a parameter of the distribution scheme.
For an initial active space where $N = L$, the number of calculations with a small active space $N^\mathrm{small}_\mathrm{calc}$ scales as
\begin{align}
N^\mathrm{small}_\mathrm{calc} = \frac{L^2}{L_\mathrm{max}^2}\, .
\label{scal}
\end{align}

\begin{figure}[h!]
\begin{center}
\includegraphics[width=0.75\textwidth]{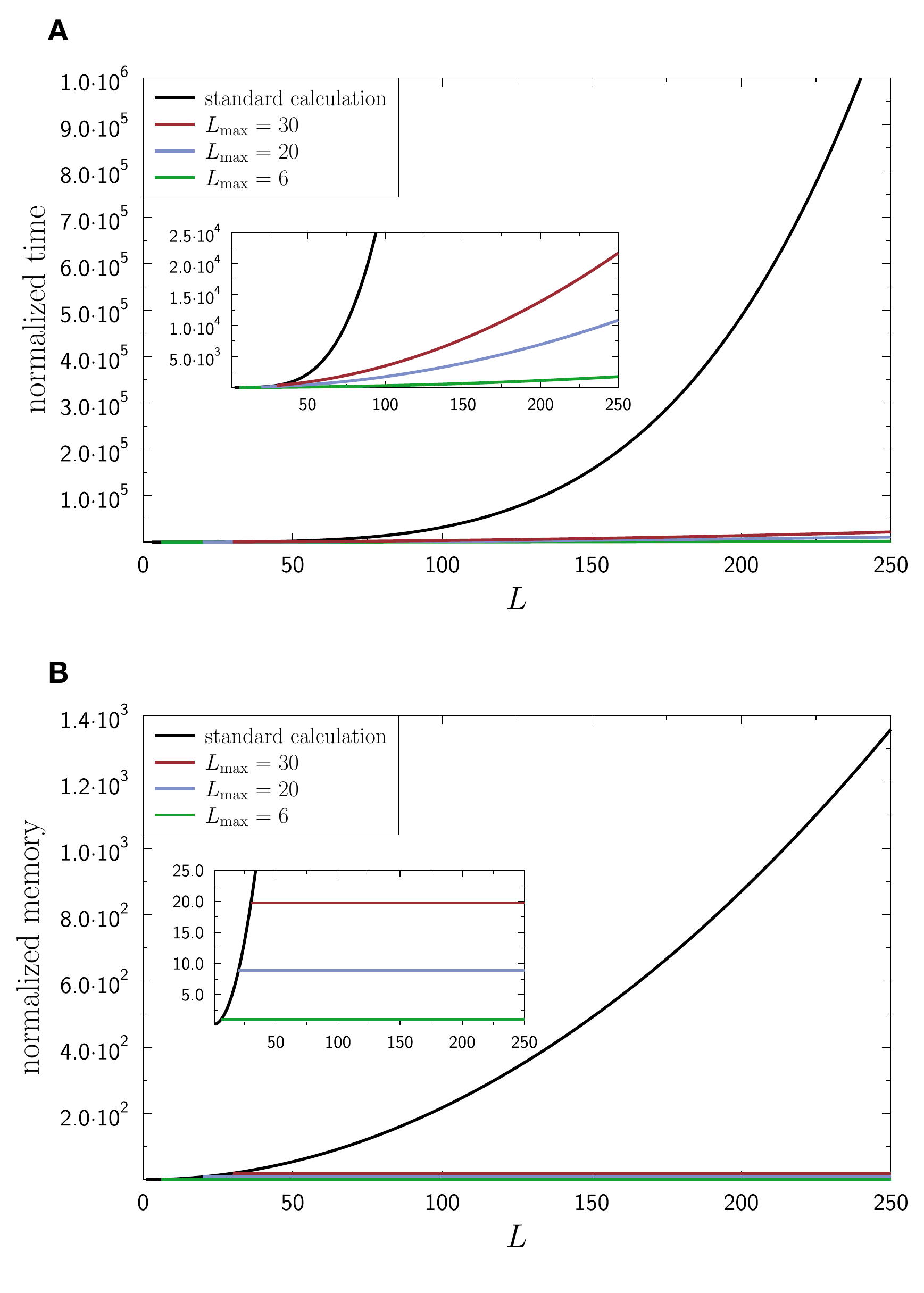}
\caption{Scaling of DMRG with respect to the number of orbitals in the active space $L$ for a fixed number of renormalized block states. The scaling is normalized to an active space with six orbitals. Panel A shows the time scaling which is proportional to $L^4$.  The black line corresponds to the standard algorithm, whereas the red, blue, and green lines correspond to the distributed variant with $L_\mathrm{max} = 6, 20$\,, and $30$, respectively. Panel B shows the memory scaling which is proportional to $L^2$. The lines for the distributed variant are horizontal because only the memory requirement for a single calculation is of interest here.}
\label{scaling}
\end{center}
\end{figure}

Without inclusion of spin or point-group symmetry the time required for a DMRG calculation scales roughly with $L^4$, whereas the memory requirement scales with $L^2$.\cite{wout14}
Fig.~\ref{scaling} shows the scaling of the standard full-size calculation compared to the distribution scheme for several choices of $L_\mathrm{max}$ normalized to the cost of a CAS(6,6) calculation.
In the inset of panel A of that figure it is demonstrated that the scaling of the computational time for the distribution scheme is controlled by the number of individual small CAS calculations ($L^2$-scaling, see Eq.~(\ref{scal})) with a prefactor given by the cost of the individual calculation.
In the right panel, the scaling of the memory requirement is displayed.
There is no scaling with the number of orbitals in the case of the distribution scheme because only the memory requirement of the individual calculation limits the feasibility of a calculation.

While this distribution scheme radically reduces the scaling and overall computational cost of the entanglement entropy calculation, it is important to note that the resulting entanglement entropies should not be mistaken for the correct ones obtained from a converged calculation including all orbitals.
The scheme does only guarantee that each occupied orbital is combined with each virtual orbital but does not guarantee all possible combinations of orbitals.
Therefore, correlation effects between more than two orbitals are not fully accounted for.

We additionally investigated a random selection of orbitals for the small CAS subspaces of size $L_\mathrm{max}$.
For a sufficiently large number of randomly chosen subspace calculations, the orbital entanglement entropies will converge.
In this random approach, the description of correlation between more than two orbitals is improved by including more
orbital combinations in an increasing number of exploratory calculations.
While this is certainly a desired property, the number of small CAS calculations required to converge the orbital entanglement entropies easily becomes computationally too demanding.
We therefore decided to apply only the distribution scheme depicted in Fig.~\ref{distribute} in this work 
but note that other schemes could be employed as well. E.g., one may combine the two schemes described above so that the randomly chosen subspace calculations add information about correlation between more than two orbitals to the occupied/virtual orbital pair correlation.
We anticipate that such a hybrid approach will be of particular interest for molecules where a simple bonding/anti-bonding orbital-pair structure does no longer exist (as in open-shell transition-metal complexes and clusters).

As a proof of principle we investigated the distribution scheme in a calculation on buckminsterfullerene C$_{60}$.
The valence orbital space of this molecule comprises 240 orbitals out of which 60 orbitals form the $\pi$-system and are weakly statically but not strongly correlated.\cite{lee19}
Our scheme should reliably identify these 60 orbitals.
For reasons mentioned above we calculated the entanglement entropies in a localized orbital basis obtained from a Pipek-Mezey\cite{pipe89} localization procedure.
We chose $L_\mathrm{max} = 20$ and calculated the entanglement entropies for 144 CAS(20,20) active spaces.
\begin{figure}[t!]
\begin{center}
\includegraphics[width=\textwidth]{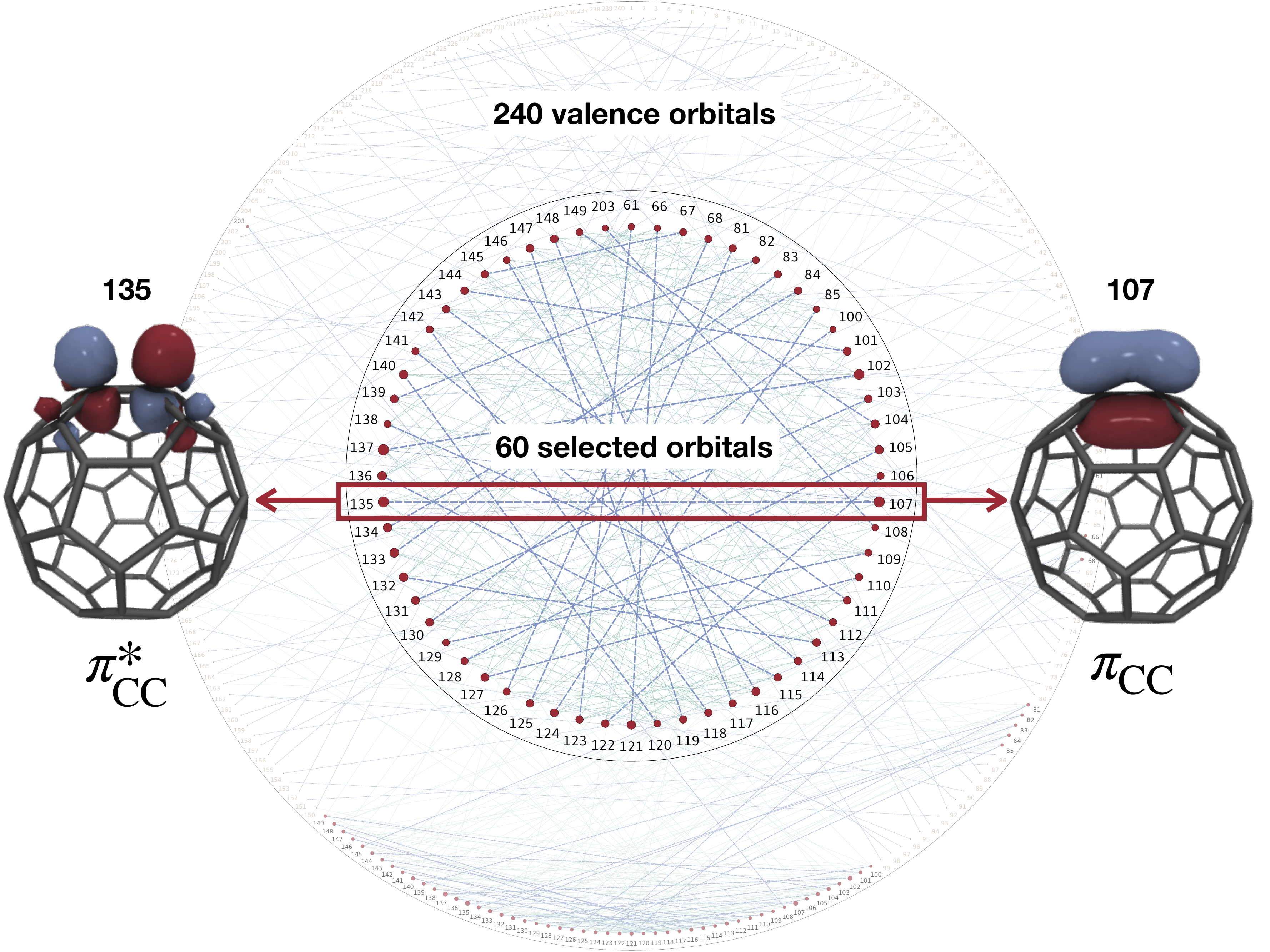}
\caption{Entanglement diagram for all 240 valence orbitals of C$_{60}$ in the background, whereas the reduced entanglement diagram for the 60 automatically selected orbitals (the full $\pi$-space) is shown in the center. In this localized basis the entanglement diagram shows a formation of orbital pairs connected by strong mutual information and these orbitals can be identified as $\pi/\pi^*$ orbital pairs. An example of such an orbital pair is shown on the left ($\pi^*$, orbital no. 135) and right hand side ($\pi$, orbital no. 107) of the figure.}
\label{result_fullerene}
\end{center}
\end{figure}
The results are summarized in Fig.~\ref{result_fullerene}. The combined entanglement diagram is displayed in the background but the circular diagrams become hardly readable for active spaces of this size.
The automated active space selection applied to the single-orbital entropies obtained with the distribution scheme leads to the 60 $\pi/\pi^*$-orbitals exactly as expected.
A closer inspection shows that 30 $\pi/\pi^*$-pairs can be identified that are connected by a strong mutual information element and an example of such an orbital pair is displayed in the same figure.

The result for C$_{60}$ is very encouraging and indicates that the distribution scheme is capable to reliably identify strongly correlated orbitals for large molecules with many valence orbitals.
In the future, we will further study this feature of \textsc{autoCAS} on different classes of compounds  such as polynuclear transition metal complexes.\cite{sinh17}
We note, however, that this approach cannot be applied for excited states if the entanglement pattern is very different from the ground state.
This is because in DMRG excited states are calculated by a ground-state optimization in the subspace orthogonal to already optimized lower-lying states.
Based on the notion that optimal active spaces of ground and excited states usually vary in at most a few orbitals, the active space of the ground state can serve as a \textit{base active space} in such a case.
A suitable active space for the excited state can then be automatically determined by stepwise inclusion of inactive valence orbitals to the base active space and identifying highly correlated orbitals in the same manner as described above for the distribution scheme.
This approach, however, is rather expensive for large base active spaces because it requires two DMRG calculations (one for the ground and one for the excited state) for each of the incrementally extended active spaces.

\section{Recommended settings}
\label{recommended}

Since \textsc{autoCAS} is designed to fully automate (or substantially support) multi-configurational calculations for a wide range of applications, the template files can be adjusted with a large degree of flexibility.
While this is obviously intended, it is desirable to have a set of recommended settings that have been shown to give convincing results.
These settings will mostly be the default options of the fully automated mode of \textsc{autoCAS}.
In this section we summarize the most important of these settings.

An exceedingly large part of static correlation can be described in a minimal basis unless highly excited (Rydberg) states are of interest.
In order to save computational time during the integral evaluation and to have chemically interpretable orbitals, we recommend to carry out the automated active space selection in a minimal basis.
If $3d$ transition metal atoms are present, a second $d$-shell has to be added to account for the double-shell effect.
Once the active space is automatically selected, the optimized orbitals may be projected onto a larger basis with the EXPBAS subroutine in \textsc{openMolcas}, which only requires the orbital file of the optimized orbitals and the one- and two-electron integrals in the small and large atomic-orbital basis.
The projected orbitals must then be reoptimized before dynamical correlation is calculated by means of a CASPT2 or NEVPT2 calculation.
The reoptimization of the orbitals usually requires less than ten iterations and hence justifies the initial choice of a minimal basis set.
In \textsc{openMolcas}, the projection onto a larger basis is possible only if the basis set is of the same type.
We therefore recommend the ANO-RCC family of basis sets\cite{roos04,widm90} that has been tailored to yield reliable results in multi-configurational calculations.

The interpretation of the active space selection is greatly simplified in a localized basis and this is therefore a suitable choice.
In addition, such a basis complies with the often rather local character of the electron correlation as briefly discussed in the previous section.
In our workflow, we hence sequentially localize the occupied and virtual subset of the initial Hartree--Fock orbitals with the Pipek-Mezey scheme.
As the optimized CASSCF or DMRG-SCF orbitals are usually of local character as well, this additional step further accelerates the convergence in the final orbital optimization step.

The degree of static correlation is dependent on the molecular orbital basis (we discussed this recently in detail on the example of H$_2$ in Ref.~\citenum{stei17a}) and it is hence possible that the final orbital optimization has a strong effect on the orbital entanglement.
We therefore recommend to always reiterate the automated active space selection in the optimized orbital basis.
In all our studies so far, however, we found that at most one reiteration is required to converge the active space.

Finally, we note that DMRG is prone to get stuck in local minima during the wave function optimization.
This can be avoided by a suitable initial guess and an appropriate ordering of the orbitals.
Although we did not find the partially converged calculations to be especially prone to this phenomenon, we implemented an optimized Fiedler ordering\cite{fied73,fied75,barc11} based on an initial mutual information and the CI-DEAS initial guess\cite{lege03,barc11} in the \textsc{openMolcas}/\textsc{QCMaquis} interface.
In order to assess the convergence of the entanglement measures, the calculations should be repeated with two different values for the number of renormalized block states $m$.

\section{Perspective and outlook}

Currently, \textsc{autoCAS} is implemented as a graphical user interface (GUI) to minimize manual work, provide an easy way to modify the calculation, and constitute a low entry-barrier for scientists that are new to the field of multi-configurational calculations.
If the automated active space selection is to be integrated as one step out of many in a complex protocol, a GUI will be unsuitable.
Examples for such protocols are automated structure explorations,\cite{simm19} automated benchmarking or a recently proposed high-throughput protocol for computational photobiology.\cite{mela16}
For these cases, a command line version of \textsc{autoCAS} will be beneficial.
Due to the object-oriented design of \textsc{autoCAS} such an adaptation can be realized without changing the key elements of the program and will be provided in the near future.

The orbital localization with the Pipek-Mezey scheme is very simple and efficient especially in a minimal basis.
A closely related class of localized orbitals are Knizia's intrinsic bond orbitals\cite{kniz13} that are formulated in a polarized minimal basis and the localization functional is the same as in the Pipek-Mezey scheme with an exponent of four.
As these orbitals are simple to implement, they correspond to chemical bonds and the localization itself is stable, they provide an adequate alternative to our current localization scheme and might be considered in the future.

The possibility to interface other electronic-structure programs will be another valuable addition to the existing functionality.
Such an extension is, although in principle simple, complicated in practice because of the heterogenous file and input formats employed by the different programs available.
As mentioned above, however, the automated selection itself may already now be carried out based on entanglement entropies obtained from arbitrary multi-configurational calculations by manually manipulating the \textsc{autoCAS} save files.

An additional highly desirable feature is a fully automated protocol for multi-configurational calculations along a reaction coordinate as described in Ref.~\citenum{stei17}.
This would include the propagation of orbitals optimized for similar structures along the predefined reaction coordinate and an automated reoptimization of previously optimized orbitals if the automatically selected active spaces had increased along the reaction coordinate.
At this point, \textsc{autoCAS} already greatly simplifies such a calculation as it allows arbitrary initial orbitals for a given structure so that the calculation for each structure is automated.
A fully automated procedure for the propagation of optimized orbitals along a reaction coordinate, however, is left for future work.

\section{Conclusions}

We presented details on the implementation of our \textsc{autoCAS} program for fully automated multi-configurational calculations.
Currently, \textsc{autoCAS} presents a GUI interfacing the \textsc{openMolcas} electronic structure program and the \textsc{QCMaquis} DMRG program.
Multi-configurational calculations can be carried out fully automatically with minimal manual input such as basis set, molecular charge, and spin state of the molecule.
The automated active space selection protocol of Ref.~\citenum{stei16} defines the core of the program.
These calculations require not more manual input than DFT calculations that are carried out routinely by non-experts.
A second important feature of \textsc{autoCAS} is the automatic evaluation of the $Z_{s(1)}$ multi-configurational diagnostic that signals whether a multi-configurational calculation is required at all or whether an overall more accurate single-configurational method such as CCSD(T)-F12\cite{tenn04,adle07,kniz09} or its linear-scaling variants\cite{guo18,ma18a,ma18} can be applied.

We introduced an extension of the original orbital selection protocol implemented in \textsc{autoCAS} that represents a distribution scheme for the entanglement analysis and automated active space selection for molecules with very large valence spaces.
We then discussed the scaling and limitations of this scheme and demonstrated our approach for the active space selection of buckminsterfullerene C$_{60}$. 
The application of this distribution scheme to other classes of molecules as well as a command line version and the interfacing with other electronic-structure programs is left for future work.
We note, however, that the distribution scheme can also be beneficially employed for small valence orbital spaces if the exploratory DMRG
calculation requires memory resources on computer hardware that are not accessible due to local computing limitations. 

\textsc{autoCAS} is part of the SCINE project for chemical interaction networks and can be downloaded free of charge as a binary file for UNIX operating systems.\cite{autocas}


\providecommand{\latin}[1]{#1}
\makeatletter
\providecommand{\doi}
  {\begingroup\let\do\@makeother\dospecials
  \catcode`\{=1 \catcode`\}=2 \doi@aux}
\providecommand{\doi@aux}[1]{\endgroup\texttt{#1}}
\makeatother
\providecommand*\mcitethebibliography{\thebibliography}
\csname @ifundefined\endcsname{endmcitethebibliography}
  {\let\endmcitethebibliography\endthebibliography}{}


\end{document}